\documentclass[showpacs,preprint,aps]{revtex4}
\usepackage{rotating}
\usepackage{graphicx}
\usepackage{dcolumn}
\usepackage{bm}
\usepackage{epsfig}
\usepackage{amsmath}
\usepackage{subfigure}

\def\be{\begin{equation}}
\def\lan{\left\langle}
\def\ran{\right\rangle}
\def\ee{\end{equation}}
\def\barr{\begin{array}}
\def\earr{\end{array}}

\def\l{\left}
\def\r{\right}
\def\dis{\displaystyle}
\def\ed{\end{document}}

\def\co{{\cal O}}
\def\ch{{\cal H}}

\begin{document}
\title{Deformed shell model results for neutrinoless double beta 
decay of nuclei in A=60-90 region}
\author{R. Sahu$^1$}
\email{rankasahu@rediffmail.com}
\author{V.K.B. Kota$^{2,3}$}
\affiliation{$^1$Physics Department, Berhampur University, Berhampur-760 007,
Odisha, INDIA}
\affiliation{$^2$Physical Research Laboratory, Ahmedabad - 380 009, INDIA \\
$^3$Department of Physics, Laurentian  University, Sudbury, ON P3E
2C6, CANADA }
\begin{abstract}

Nuclear transition matrix elements (NTME) for the neutrinoless double beta
decay   of $^{70}$Zn, $^{80}$Se and $^{82}$Se nuclei are calculated within the 
framework of the deformed shell model based on Hartree-Fock states. For
$^{70}$Zn, jj44b interaction in  $^{2}p_{3/2}$, $^{1}f_{5/2}$,  $^{2}p_{1/2}$
and  $^{1}g_{9/2}$ space  with $^{56}$Ni as the core is employed. However, for 
$^{80}$Se and $^{82}$Se nuclei, a modified  Kuo  interaction with the above core
and model space are employed. Most of our calculations in this region were
performed with this effective interaction. However, jj44b interaction has been
found to be better for $^{70}$Zn. The above model space was used in many
recent shell model and interacting boson model calculations for nuclei in this
region. After ensuring that DSM gives good description
of the spectroscopic properties of low-lying levels in these three nuclei
considered, the NTME are calculated. The deduced half-lives with these NTME,
assuming neutrino mass is 1 eV,  are  $1.1 \times 10^{26}$ yr, $2.3 \times
10^{27}$ yr and $2.2 \times 10^{24}$ yr for  $^{70}$Zn, $^{80}$Se and $^{82}$Se,
respectively.

\end{abstract}

\pacs{23.40.Hc, 21.10.Tg,  21.60.Jz, 27.50.+e}

\keywords{neutrinoless double beta decay, positron modes, deformed shell model, 
A=60-90 nuclei}

\maketitle

\section{Introduction}

Neutrinoless double beta decay ($0\nu \beta \beta$ or $0\nu\,DBD$) which
involves emission of two electrons without the accompanying neutrinos and which
violates  lepton number conservation has been an important and challenging
problem both  for the experimentalists and theoreticians.  Recent neutrino
oscillation  experiments have demonstrated that neutrinos have mass.  The
observation of $0\nu \beta \beta$ decay is expected to provide information
regarding  the absolute  neutrino mass which is not known. To extract neutrino
mass, the nuclear matrix elements must be known from a reliable nuclear model
and hence the main goal of nuclear theorists is to calculate the nuclear
transition matrix elements as  accurately as possible. On the other hand
experimental programmes have  been initiated at different laboratories across
the globe to observe this  decay and the experiments are already in advanced
stages of development. The most recent experimental results of  $0\nu \beta
\beta$ decay of $^{136}$Xe have been reported by KamLand-Zen  collaboration
\cite{Kam} and EXO 200 collaboration \cite{Exo} and they  give a lower limit of
$3.4\times 10^{25}$ yr for the half-life.  On the other hand, phase I results
from GERDA experiment \cite{Gerda} for $^{76}$Ge are published recently
giving a lower limit of $3.0\times 10^{25}$ yr for the half-life. 

Nuclear transition matrix elements (NTME) are the essential ingredient for
extracting the neutrino mass from the half lives. There has been considerable
effort to obtain NTME for various candidate nuclei and they have been
calculated  theoretically using a variety of nuclear models: (i) large scale
shell model (SM) \cite{SM}; (ii) quasi-particle random phase approximation
(QRPA) and its variants \cite{QRPA1,QRPA2,QRPA3};  (iii) proton-neutron
interacting boson model (IBM-2) \cite{IBM1}; (iv) particle number and angular
momentum projection including configuration mixing within the generating
coordinate method framework (GCM+PNAMP) \cite{DFT}; (v) projected
Hartree-Fock-Bogoliubov (PHFB) method with pairing plus  quadrupole quadrupole
interaction \cite{HFB}. Detailed comparative study of  the results from these
methods is discussed recently in \cite{IBM2,JS1}. 

Besides the two electron mode, it is also possible to have neutrinoless positron
double beta decay and this can come in three modes:  (i) double $\beta^+$
($\beta^+\beta^+$), (ii) $\beta^+$ and electron capture  ($\beta^+$EC) and (iii)
double electron capture (ECEC) . The later process can proceed only by emission
of extra particles or as a resonant process \cite{JS2}.
  All these three modes combined are referred to
as $0\nu\; e^+$DBD. There are now efforts to observe  $0\nu$ (and also $2\nu$)
$e^+$DBD in some nuclei; see for example  \cite{zn64,se74,sr84}.

Over the last many years, the Deformed Shell Model (DSM)  (based on Hartree-Fock
states) has been used to study various properties of nuclei in the mass A=60-90
region (also in A=44-60) with reasonable success. They include: (i)
spectroscopic properties such as band structures, shapes and shape coexistence,
nature of band crossings, electromagnetic transition probabilities
\cite{sahu1,sahu2,sahu3,br8082,npa96,sahu4}; (ii) $T=1$ and  $T=0$ bands in N=Z
odd-odd nuclei and $T=1/2$ bands in odd-A nuclei by including isospin projection
\cite{sk1,sk2,msk};  (iii)  transition matrix elements for $\mu-e$ conversion in
$^{72}$Ge  \cite{app1} and in the analysis of data for inelastic scattering of
electrons from $fp$-shell nuclei \cite{app3}. More importantly, this model has
also been used for studying $2\nu$ double beta decay, in a first attempt, for
$^{76}$Ge $\rightarrow$ $^{76}$Se in \cite{ge76} with  reasonable success.
The DSM results for the spectroscopy of $^{62}$Ga have recently been compared
with the predictions of spherical shell model \cite{ga62}. The $T=0$ and $T=1$
energy levels and the isoscalar and isovector electromagnetic
transition probabilities generated in both these models are found to be similar.
Following this and the success of DSM in explaining spectroscopic properties of 
nuclei in A=60-90 region, we have started a study of DBD in A=60-90 nuclei.

As shown in table \ref{tab:table1}, there are eight candidate nuclei in the
A=60-90 region with $30 \leq \mbox{Z} \leq 40$ and $\mbox{N} \leq 48$ which can
undergo double beta decay. We have already applied DSM to study half lives for
$2\nu\;e^+$DBD in $^{78}$Kr \cite{SK-kr}, in $^{74}$Se \cite{SK-se}, in
$^{84}$Sr \cite{SK-sr} and in $^{64}$Zn \cite{SSK}. Going further, more recently
we have also calculated NTME for neutrinoless positron double  beta decay $0\nu
\beta^+ \beta^+$ and $0\nu \beta^+$EC for all the  above four nuclei \cite{SSK}.
We ensured that DSM gives a good description of  the spectroscopic properties
for low-lying states for these nuclei.  The deduced half-lives with these NTME,
assuming neutrino mass is 1 eV,  are  smallest for $^{78}$Kr with half life for
$\beta^+$EC decay being $\sim 10^{27}$ yr. For all others, the half lives are in
the range of $\sim 10^{28}$ to $10^{29}$ yr. 

\begin{table}[h]
\caption{\label{tab:table1}DBD candidates with A=60-90, 30$\leq$Z$\leq$40
and N $\leq 48$.}
\begin{tabular}{cc}
\hline
$(0+2\nu)\beta^-\beta^-$~~~ &~~~ $(0+2\nu)e^+$DBD\\
\hline
$^{70}$Zn & $ ^{64}$Zn\\
$^{76}$Ge & $ ^{74}$Se\\
$^{80}$Se & $ ^{78}$Kr\\
$^{82}$Se & $ ^{84}$Sr\\
\hline
\end{tabular}
\end{table}

In addition to completing the study of the four candidate nuclei listed in Table
I that undergo $e^+$DBD, in \cite{SSK} we have also reported DSM results for
$0\nu\;$DBD for $^{76}$Ge. With this, we are left with the analysis of $2\nu$
and $0\nu$ DBD in $^{70}$Zn, $^{80}$Se and $^{82}$Se nuclei (first DSM results
for $2\nu$ DBD for $^{82}$Se were given in \cite{SSK-2}). Our purpose in the
present paper is to present DSM results for these nuclei.  We will give a 
preview.

In Section 2 discussed briefly are the formula for half life for $0\nu$DBD, the
transition operator generating NTME and the DSM model. Section 3 gives DSM
results for $^{70}$Zn first for spectroscopic properties and then the results
for both $2\nu$ and $0\nu$ double beta decay half lives. In Section 4, DSM
results for $2\nu$ and $0\nu$  double beta decay half lives for $^{80}$Se are
given and similarly in Section 5 for $^{82}$Se.  Not many theoretical
calculations of  spectroscopic properties of $^{70}$Zn are available in the
literature.  Our results obtained using DSM are presented in Section 3. Also,
DSM results for orbit occupancies for all  the three nuclei are discussed in
Sections 3-5. Finally, Section 6 gives  conclusions. 

\section{Formalism}

\subsection{$0\nu$DBD half life and transition operator}

Half-life for  $0\nu$DBD for the 0$^+_i$ ground state (gs) of a initial
even-even nucleus decaying to the 0$^+_f$ gs of the final even-even nucleus is
given by \cite{IBM2}
\be
\l[ T_{1/2}^{k: 0\nu}(0^+_i \to 0^+_f) \r]^{-1} =  G^{0\nu}(k)\; \l(g_A\r)^4
\l| M^{0\nu} (0^+)\r|^2 \l(\dis\frac{\lan m_\nu \ran}{m_e}\r)^2 \;,
\label{eq.dbd1}
\ee
where $\lan m_\nu \ran$ is the effective neutrino mass (a combination of
neutrino mass eigenvalues and it also involves neutrino mixing matrix) and $k$
denotes the decay mechanism (light neutrino exchange, heavy neutrino exchange
and Majoron emission - in the present paper light neutrino exchange is
assumed).  The $G^{0\nu}(k)$ is phase space integral 
(kinematical factor) dependent on charge, mass and available energy for the
$0\nu$DBD process.   In Eq. (\ref{eq.dbd1}), the $M^{0\nu}$ is the nuclear
transition matrix element (NTME) of the $0\nu$DBD transition operator and it is
a sum of a Gamow-Teller like ($M_{GT}$), Fermi like ($M_F$) and tensor ($M_T$)
two-body operators. As it is well known that the tensor part contributes only up
to 10\% of the matrix elements \cite{IBM2},  we will neglect the tensor part.
Then we have, from the closure approximation which is well justified for
$0\nu$DBD,
\be
\barr{rcl}
M^{0\nu} (0^+) & = & M^{0\nu}_{GT} (0^+) - \dis\frac{g_V^2}{g_A^2} 
M^{0\nu}_{F} (0^+) = \lan 0^+_f \mid\mid \co(2:0\nu) \mid\mid 0^+_i \ran \;,
\\
\co(2:0\nu) & = & \dis\sum_{a,b} \ch(r_{ab}, \overline{E}) 
\tau_a^+ \tau_b^+ \l( \sigma_a \cdot \sigma_b- \dis\frac{g_V^2}{g_A^2} \r)
\;.
\earr \label{eq.dbd2}
\ee
Note that $\tau^+$ changes a neutron into a proton. As seen from Eq.
(\ref{eq.dbd2}),  $0\nu$DBD half-lives are generated by the two-body transition
operator $\co(2:0\nu)$; note that $a,b$ label  nucleons. The $g_A$ and $g_V$ are
the weak axial-vector and vector coupling constants.  The $\ch(r_{ab},
\overline{E})$ in Eq. (\ref{eq.dbd2}) is the `neutrino potential'. Here
$\overline{E}$ is the average energy of the virtual intermediate states used in
the closure approximation  \cite{IBM2,IBM3,En-88,To-91,Si-09}. The form given by
Eq. (\ref{eq.dbd2}) is justified {\it only if the exchange of the light Majorana
neutrino is indeed the mechanism responsible for the $0\nu$DBD}.  The effects of
short-range correlations in the wave functions are taken into  account by
multiplying the wave function by the Jastrow function $[1 - \gamma_3e^{-\gamma_1
r_{ab}^2} ( 1 - \gamma_2 r_{ab}^2 )]$ \cite{En-88,Si-09}. Now keeping the wave
functions unaltered, the Jastrow  function can be incorporated into $\ch(r_{ab},
\overline{E})$ giving an effective $\ch_{eff}(r_{ab}, \overline{E})$,
\be
\ch(r_{ab}, \overline{E}) \to 
\ch_{eff}(r_{ab}, \overline{E}) = C_0 \times \ch(r_{ab}, \overline{E}) 
[ 1 - \gamma_3 \;e^{-\gamma_1 \; r_{ab}^2} ( 1 - \gamma_2 \; r_{ab}^2 ) 
]^2 \;.
\label{eq.dbd3b}
\ee 
The values chosen for the parameters $\gamma_1$, $\gamma_2$ and  $\gamma_3$ are
given below. The value of the constant $C_0$ is found to be 3 by comparing the
calculated half-lives and the shell model value for half lives for $^{76}$Ge 
as has been discussed in detail in \cite{SSK}. This constant is used to 
renormalize the $0\nu$ nuclear matrix elements. As seen ahead, the model 
reproduces reasonably well the spectroscopy of the parent and daughter nuclei. 
It also describes $2\nu$ half lives. But the model under estimates the
spherical shell model calculated nuclear matrix elements by a factor of ~3. 
The reasons can be two fold. One possibility is the limited number of 
configurations
taken in our calculation unlike shell model. Another plausible reason is
that shell model uses different effective interaction and these effective 
interactions are not available in the literature. As shown in \cite{IBM2},
the nuclear matrix elements in different models already differ by as much
as by a factor of 3. Hence essential physics will not be lost if we renormalize
the nuclear matrix elements by a scale factor of 3. 

There are a number of parameters in the $0\nu$DBD transition operator and the
choices made for the various parameters are: (i) $R = 1.2 A^{1/3}$ fm; (ii)
$b=1.003A^{1/6}$ fm \cite{IBM2};  (iii) $\overline{E}=1.12A^{1/2}$ MeV
\cite{To-91};  (iv)  $g_A/g_V=1.254$; (v) ($\gamma_1$, $\gamma_2$,
$\gamma_3$) in Eq. (\ref{eq.dbd3b}) are $(1.1,0.68,1)$ [Miller-Spencer]
\cite{Ho-10,Si-09}.  

There are other methods, besides using Jastrow function as in Eq. (3),  
to take into account the short range correlations.
The unitary correlation operator method (UCOM) \cite{Kort-07} recently
developed by Kortelainen et al. is one such effort. It has been shown that 
the Jastrow method exaggerates the effects of short-range correlations.
For example in $^{48}$Ca and $^{76}$Ge, the Jastrow method has been found to
lead to reduction of 30\%-40\% in the magnitude of nuclear matrix 
elements whereas in UCOM, the reduction is 7\%-16\%. The calculation 
using this method will be carried out in a separate publication.

Now we will discuss briefly the deformed shell model
formulation for  calculating NTME.

\subsection{DSM model}

In DSM, for a given nucleus, starting with a model space consisting of a given
set of single particle (sp) orbitals and effective two-body Hamiltonian, the
lowest energy intrinsic states are obtained by solving the Hartree-Fock (HF)
single particle equation self-consistently. Excited intrinsic configurations are
obtained by making particle-hole excitations over the lowest intrinsic state. 
These intrinsic states  $\chi_K(\eta)$ do not have definite angular momenta. 
and states of good angular momentum projected from an intrinsic state
$\chi_K(\eta)$ can be written in the form
\be
\psi^J_{MK}(\eta) = \frac{2J+1}{8\pi^2\sqrt{N_{JK}}}\int d\Omega D^{J^*}_
{MK}(\Omega)R(\Omega)\l| \chi_K(\eta) \ran 
\label{eq.proj}
\ee
where $N_{JK}$ is the normalization constant given by
\be
N_{JK} =\frac{2J+1}{2} \int^\pi_0 d\beta \sin \beta d^J_{KK}(\beta)
\lan \chi_K(\eta)|e^{-i\beta J_y}|\chi_K(\eta)\ran \;.
\label{eqn2}
\ee
In Eq. (\ref{eq.proj}) $\Omega$ represents the Euler angles ($\alpha$, $\beta$,
$\gamma$), $R(\Omega)$ which is equal to exp($-i\alpha J_z$)exp($-i\beta
J_y$)exp( $-i\gamma J_z$) represents the general rotation operator.  The good
angular momentum states projected from different intrinsic states are not in
general orthogonal to each other. Hence they are orthonormalized and then band
mixing calculations are performed.  For details see \cite{SSK}.

DSM is well established to be a successful model for transitional nuclei (with
A=60-90) when sufficiently large number of intrinsic states are included in the
band mixing calculations. Performing DSM calculations for the parent and
daughter  and then using the DSM wave functions, the $\co(2: 0 \nu)$ operator
matrix elements are  calculated and the results are presented in the next three
Sections. 

\section{$^{70}$Zn results}

\subsection{Spectroscopic properties}

The double beta decay of $^{70}$Zn has recently been studied experimentally by
Belli et al \cite{zn64} (they set the lower limits to the half life for
neutrinoless DBD to be $3.2 \times 10^{19}$ yr at 90\% C.L.).  Its natural
isotopic abundance is 0.62\%. Before going to double beta decay, we  will first
study its spectroscopic properties using DSM model to test  the goodness of the
model for this nucleus.

\begin{figure}
\includegraphics[width=6in]{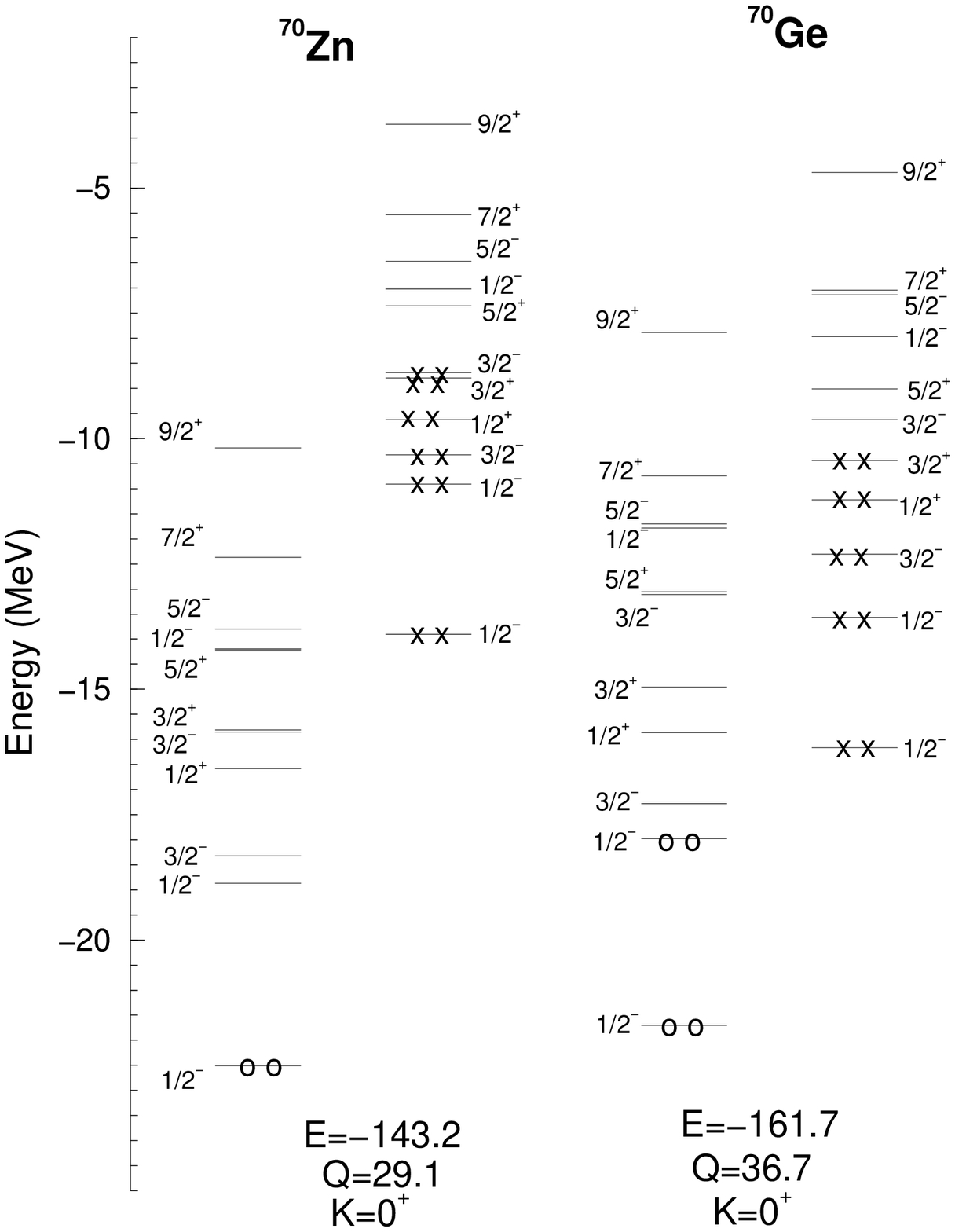}
\caption{HF single particle spectra for $^{70}$Zn and  $^{70}$Ge 
corresponding to lowest prolate configurations.  In the
figures circles represent protons and crosses represent neutrons. The
Hartree-Fock energy  ($E$) in MeV, mass quadrupole moment ($Q$) in units of
the square of the oscillator length parameter and the total $K$ quantum
number of the lowest intrinsic states are given in the figure. Each
occupied single particle orbital is two fold degenerate because of time
reversal symmetry.}
\label{fig1}
\end{figure}

$^{70}$Zn with 30 protons and 40 neutrons lie near the proton shell closure and
neutron sub-shell closure. The daughter nucleus $^{70}$Ge has two neutrons  less
and two protons more.  The spectra of ground state bands for both the nuclei are
similar. The energy levels are more or less  equispaced.  As discussed in Sect.
II.B., we first carry out an axially symmetric HF calculation for each nucleus
using the newly developed jj44b effective interaction \cite{Brown}  within the
model space consisting of the orbitals  $^{2}p_{3/2}$, $^{1}f_{5/2}$,
$^{2}p_{1/2}$ and $^1g_{9/2}$ with $^{56}$Ni  as the core. It should be
mentioned that the same model space was used in many recent shell model and
interacting boson model calculations
\cite{sm-ik1,sm-ik2,sm-ik3,sm-ik4,sm-ik5,IBM2} for $^{76}$Ge and $^{82}$Se
nuclei and $^{70}$Zn is in the same region. The spherical single particle
energies for these orbits are taken as  -9.6566, -9.2859, -8.2695 and -5.8944
MeV and are kept same both for  protons as well as neutrons. In the past we have
performed many calculations in this region using a modified Kuo interaction. 
However,  $^{70}$Zn lies very close to the proton shell closure and the modified
Kuo interaction has been found to be inadequate for this nucleus. Hence we study
this nucleus  using the jj44b effective interaction developed by Brown and his
group  \cite{Brown}.  The lowest energy HF solutions for $^{70}$Zn and its
daughter nucleus $^{70}$Ge are shown in  Fig. \ref{fig1}. The two active protons
in $^{70}$Zn occupy the lowest $k=1/2^-$ orbital. There is a well defined gap of
3.7 MeV  above the proton Fermi surface. The neutron gap above the neutron
Fermi  surface is 1.3 MeV. As a result, $^{70}$Zn has relatively stable
deformation in the ground state.  As discussed before, particle-hole excitations
over the lowest HF solution are carried out and generated a total of 226
intrinsic states (with $K=0^+$ and $K \neq 0^+$ up to 6 MeV excitation). 
Angular momentum projection from each of these intrinsic states is carried out
and then a band mixing calculation is performed.  The calculated levels are 
classified on the basis of their B(E2) values and also the structure of the
levels.

The single particle spectrum for the lowest energy HF solution for the  daughter
nucleus $^{70}$Ge is also shown in figure \ref{fig1}. The proton and neutron
gaps above their respective Fermi surfaces are less than 1 MeV. Hence one can
easily excite protons and neutrons above their Fermi surfaces. We have
considered 180 configurations with $K=0^+$ and $K \neq 0^+$ up  to 6 MeV
excitation. Good angular momentum states are projected from each of these
intrinsic states and then a band mixing calculation is performed to
orthonormalize these projected states. The calculated levels having similar
structure and connected by relatively large B(E2) are classified as belonging to
one band. 

\begin{figure}
\includegraphics[width=6in]{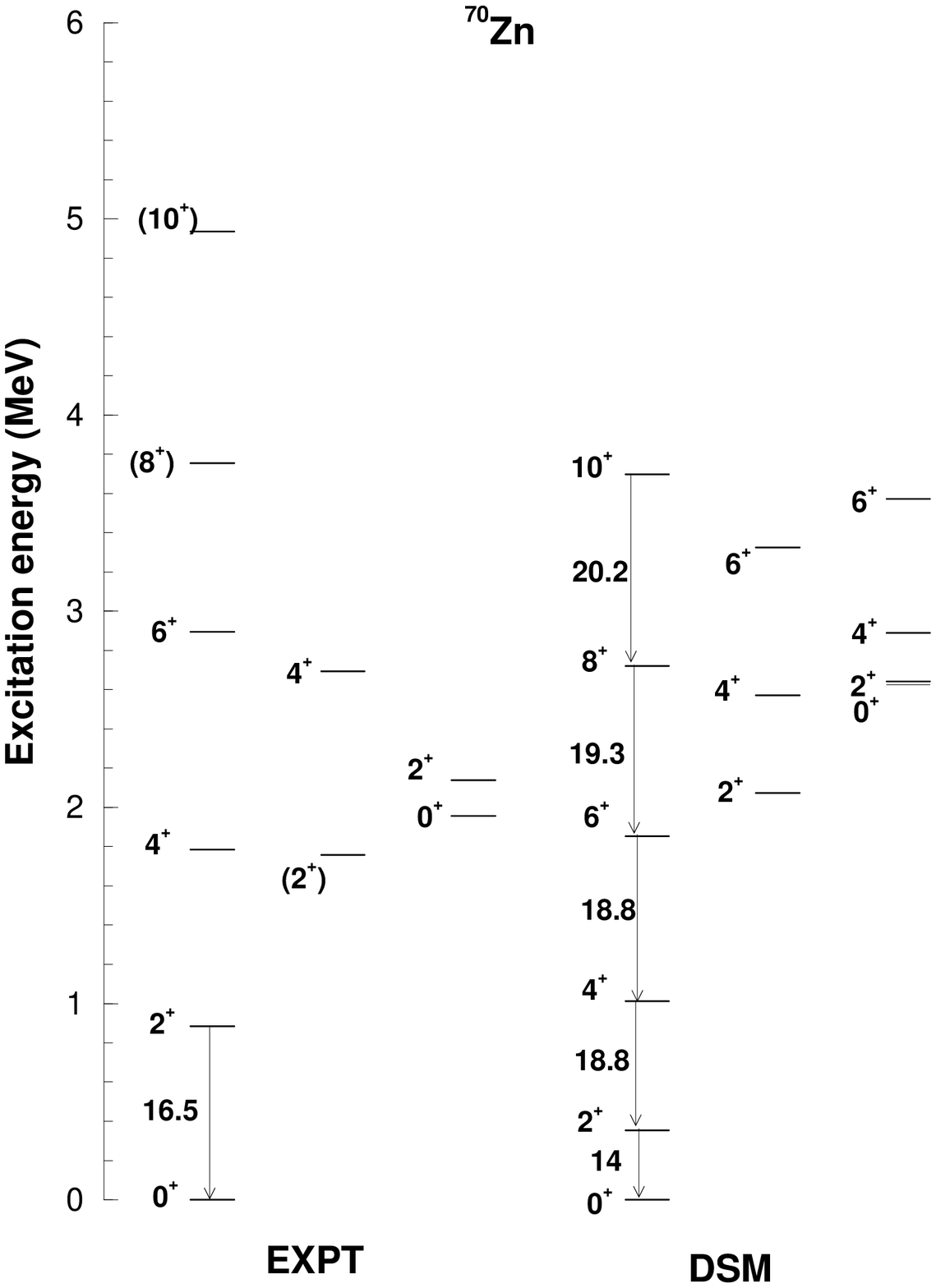}
\caption{The calculated energy levels for $^{70}$Zn are compared with 
experiment. The experimental data are from ref  \cite{nndc}. The quantities
near the arrows represent B(E2) values in W.u.
}
\label{fig2}
\end{figure}

\begin{figure}
\includegraphics[width=6in]{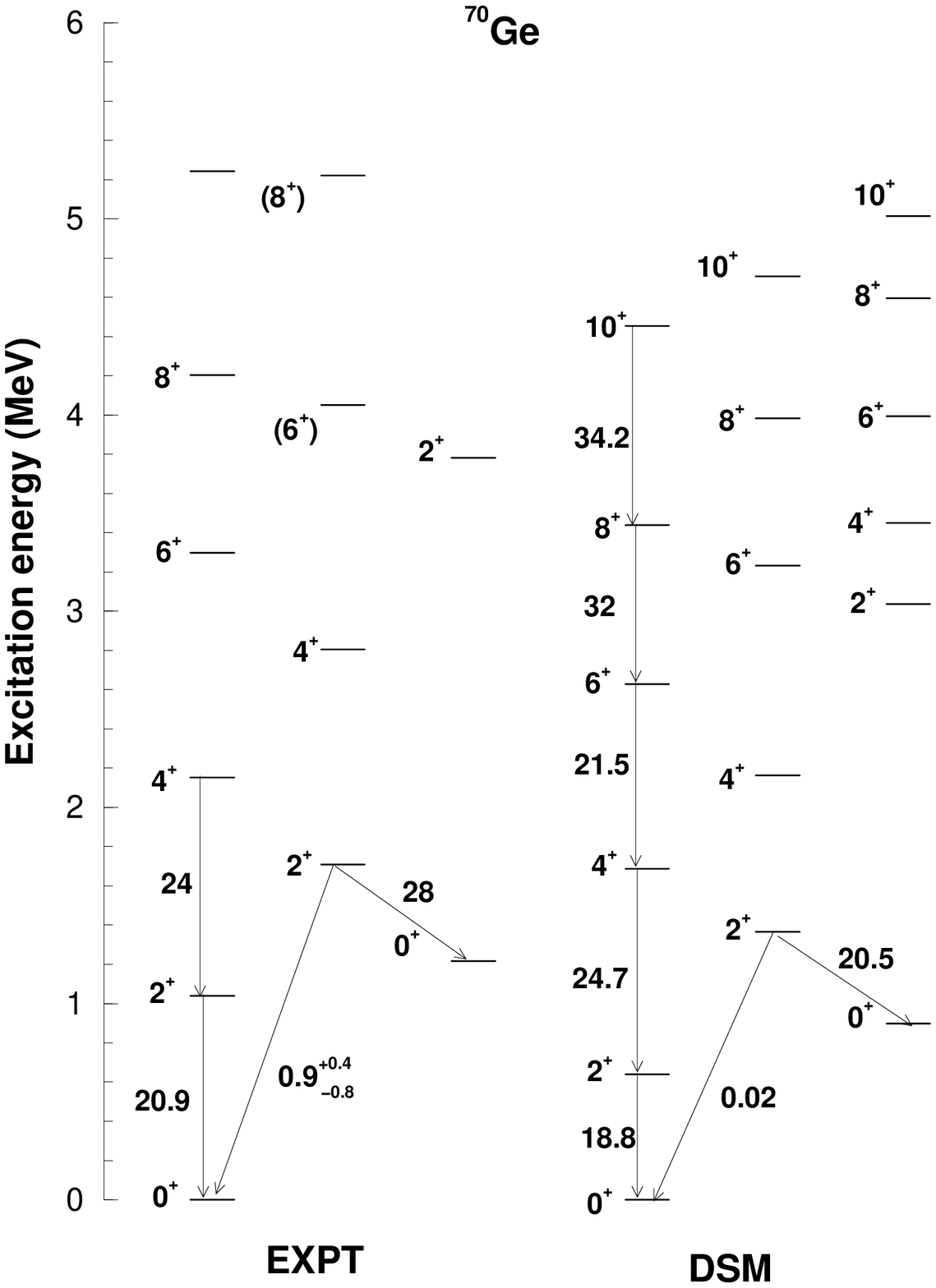}
\caption{The calculated energy levels for $^{70}$Ge are compared with 
experiment. The experimental data are from ref \cite{nndc}.
 The quantities near the arrows represent B(E2) values in W.u.
}
\label{fig3}
\end{figure}

The calculated levels and the bands for $^{70}$Zn and $^{70}$Ge are compared
with experiment in Figs. \ref{fig2} and \ref{fig3}.  The experimental data for
the two nuclei are taken from \cite{nndc}. The calculated spectrum is 
compressed compared to experiment.  Except for the $2^+ \rightarrow 0^+$
separation, the relative spacing of all the other levels are reasonable. The
ground band in both nuclei is mainly an admixture of the lowest  intrinsic
configuration at low spin. However at higher spins, there is  mixing due to
other configurations. The quasi-gamma bands in both the nuclei  are also
compared with experiment.   The quantities near the arrows represent B(E2)
values in W.u. unit. The agreement with experiment for the B(E2) values is quite
satisfactory.  The B(E2) values are calculated with  effective charges of
$e_p$=1.5 and   $e_n$=0.5. The B(E2) values provide a test of the goodness of
nuclear wave  functions generated in the model.  In view of the good agreement,
we have confidence about the suitability of the model for studying double beta
decay properties.

\begin{figure}
\includegraphics[width=6in]{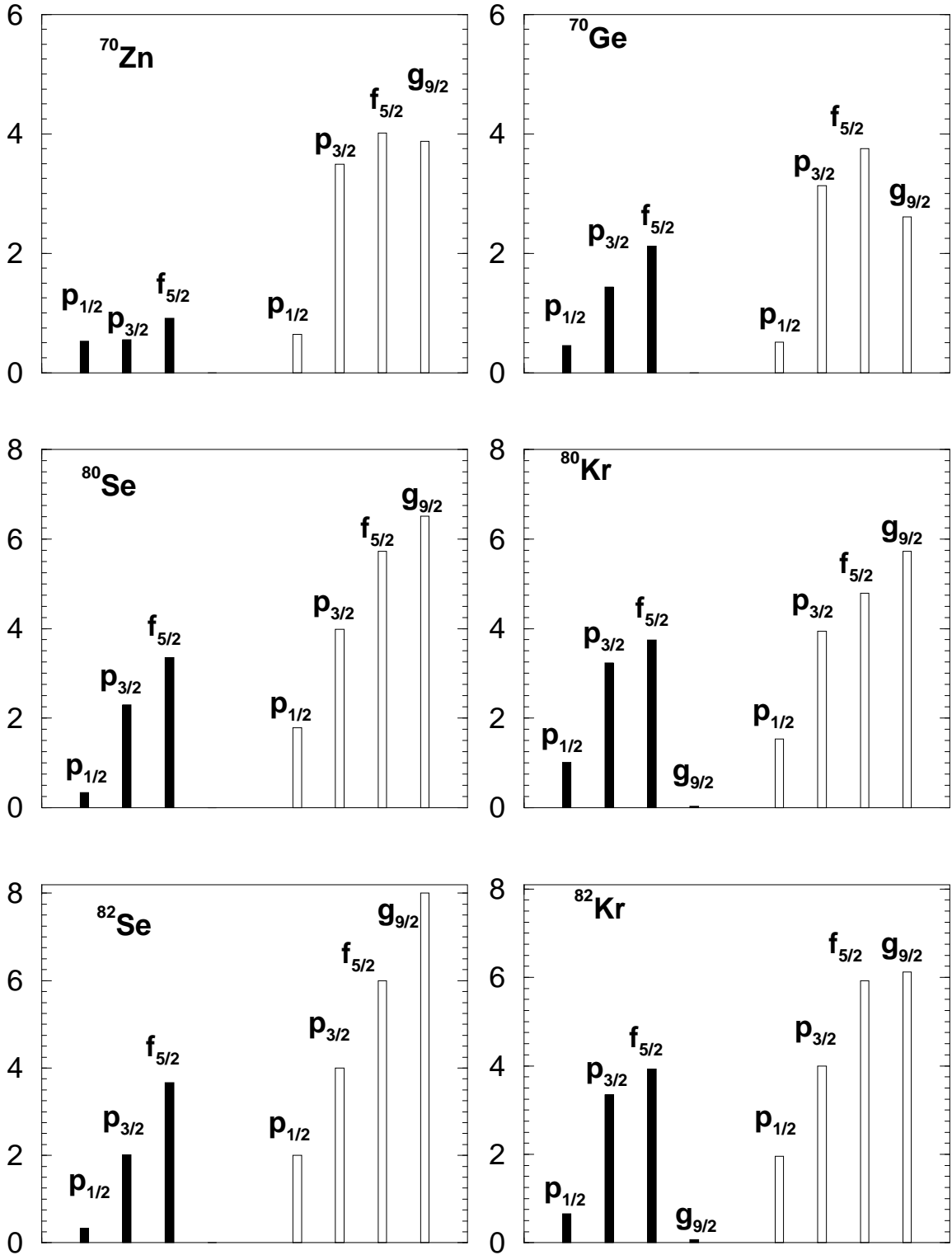}

\caption{DSM results for occupancies of different orbits. Filled bars are for 
proton  occupancies and open bars are for neutron occupancies.
}
\label{fig4}
\end{figure}

Going beyond these, orbit occupancies for protons and neutron for these  nuclei
are presented in Fig. \ref{fig4}.  It is important to add that, recently there
has been experimental efforts to measure the population of the valence orbits in
several double beta decay candidate nuclei \cite{Sch1,Ko-79,Sch2, Sch3, Sch4}. 
It will be quite useful if single nucleon transfer experiments are performed
for  $^{70}$Zn and $^{70}$Ge nuclei to test these results.

\subsection{$2\nu$ DBD half lives and $0\nu$ DBD NTME and half lives}

First we will consider $2\nu$ DBD and the  half-life for the $0_I^{+}
\rightarrow 0_F^{+}$ double beta decay  is given by \cite{Vogel}
\be
\l[ T_{1/2}^{2\nu}\r] ^{-1} = G_{2\nu } \;\l|M_{2\nu }\r|^2
\label{eqn19}
\ee
The kinematical factor $G_{2\nu} $ is independent of nuclear structure and its
value $G_{2\nu} =0.32 \times 10^{-21}$ yr$^{-1}$ 
\cite{Vogel}). On the other hand, the nuclear transition matrix elements (NTME)
$M_{2\nu}$ are nuclear model dependent and they are given by,
\begin{equation} 
M_{2\nu}=\;\dis\sum_N\;\dis\frac{\lan 0_F^+ \mid\mid
\sigma\, \tau ^{+} \mid\mid 1_N^+ \ran \lan 1_N^+ \mid\mid \sigma\,\tau
^{+} \mid\mid 0_I^+ \ran}{\l[ E_N - (E_I + E_F)/2\r]/m_e} 
\label{eqn20}
\end{equation}
where $\l| 0_{I}^{+}\ran$, $\l| 0_{F}^{+}\ran$ and  $\l| 1_{N}^{+}\ran$ are the
initial, final and virtual intermediate states respectively and $E_{N}$ are the
energies of the intermediate nucleus $^{70}$Ga. Similarly $E_I$ and  $E_F$ are
the ground state energies of the parent and daughter nuclei. We have from 
\cite{audi-1}, the atomic masses of $^{70}$Zn, $^{70}$Ga and $^{70}$Ge to be
$-69.5646$, $-68.9101$ and $-70.5631$ MeV respectively (for nuclear mass, we
need to subtract the mass of the electrons from the atomic mass).  For
$^{70}$Ga, $1^+$ is the ground state. Then, with $E_{1^+}$ denoting the relative
energies of the $1^+$ states in $^{70}$Ga with respect to the lowest $1^+$
state, we finally obtain $[ E_N - (E_I + E_F)/2 ] = [1.1537 + E_{1^+}]$ MeV. 
DSM is used to calculate $E_{1^+}$. The nuclear matrix elements given ahead
correspond to the values of $M_{2\nu}$ with $m_e$=1.

In the DSM calculations we have considered 30 lowest intrinsic states with   $K
= 0^+$ for $^{70}$Zn, 26 lowest intrinsic configurations with $K = 0^+$    for
the daughter nucleus $^{70}$Ge and 65 intrinsic states with $K = 1^+$ or  $K =
0^+$ for the intermediate nucleus $^{70}$Ga. The intrinsic states  with $K =
1^+$ or $K = 0^+$ for $^{70}$Ga are generated by making particle-hole 
excitations over the lowest HF intrinsic state generated for this nucleus. We
project out $1^+$ states from each of these intrinsic states and  then perform a
band mixing calculation as discussed above. Taking the  phase space factor $0.32
\times 10^{-21}$ yr$^{-1}$ \cite{Vogel} the DSM value for the $2\nu$ DBD
half-life is $3.39 \times 10^{23}$. Bobyk et al \cite{Bobyk} have evaluated the
half life for the $2\nu \beta\beta$  decay using different variants of QRPA with
different values of $g_{ph}$ and $g_{pp}$ and their value varies from $5 \times
10^{20}$ to $6.4 \times 10^{23}$. Our calculated half life lies near the value
in the upper limit. In our calculation, the nuclear matrix element $M_{2\nu}$
comes out to be 0.19 $MeV^{-1}$ which is smaller than the value given by 
Suhonen 
\cite{Suh-3} by a factor of around 2.4. The contributions to the nuclear matrix
element by the first two $1^+$ states of the intermediate nucleus are 0.058 and 
0.095 $MeV^{-1}$, respectively. Thus these two states contribute predominately 
to the nuclear matrix element and this is close to the single-state dominance
predicted in \cite{CS-1}.

Spectroscopic results and $2\nu$DBD half lives  show that we can use DSM for
reliable predictions for $0\nu$ DBD. Turning to this, in the calculation of
half-lives for $0\nu$ DBD we have considered 30 intrinsic states with $K = 0^+$
for $^{70}$Zn and 26 intrinsic configurations with $K = 0^+$  for the daughter
nucleus $^{70}$Ge as in $2\nu$ case. The calculated NTME for $^{70}$Zn is 1.99. 
The GT contribution is 1.748 and the Fermi part is -0.379. We see that the Fermi
matrix element is small compared to the GT matrix element and hence there is no
isospin contamination although we have used proton-neutron formulation in DSM
without isospin projection. In Table II, the DSM value for $M^{0\nu}$ is
compared with the available QRPA values \cite{Suh-3} obtained with  different
$g_A$ values and different sets of single particle energies for the orbits
chosen in the calculations. The values with $g_A=1.25$ are close to the DSM
values.   Now, using Eq. (1),  taking the neutrino mass to be 1 eV and the 
phase space factor $0.23 \times 10^{-26}$ yr$^{-1}$ from \cite{Vogel}, the
calculated  half-life from DSM for  $0\nu \beta\beta$ decay is  $1.1 \times
10^{26}$ yr (note that in \cite{Vogel}, the factors $(g_A)^4$ and $(m_e)^2$
appearing in Eq. (1) are absorbed in the phase space factor $G^{0\nu}$).  At
present very low lower bounds ($\geq 3.2 \times 10^{19}$ yr at 90\% C.L.) for
the $0\nu$ DBD half life for $^{70}$Zn is known from experiments \cite{zn64}.
The half-lives are displayed in table \ref{tab3}. 

\section{$^{80}$Se results}

\subsection{Spectroscopic properties}

\begin{figure}
\includegraphics[width=6in]{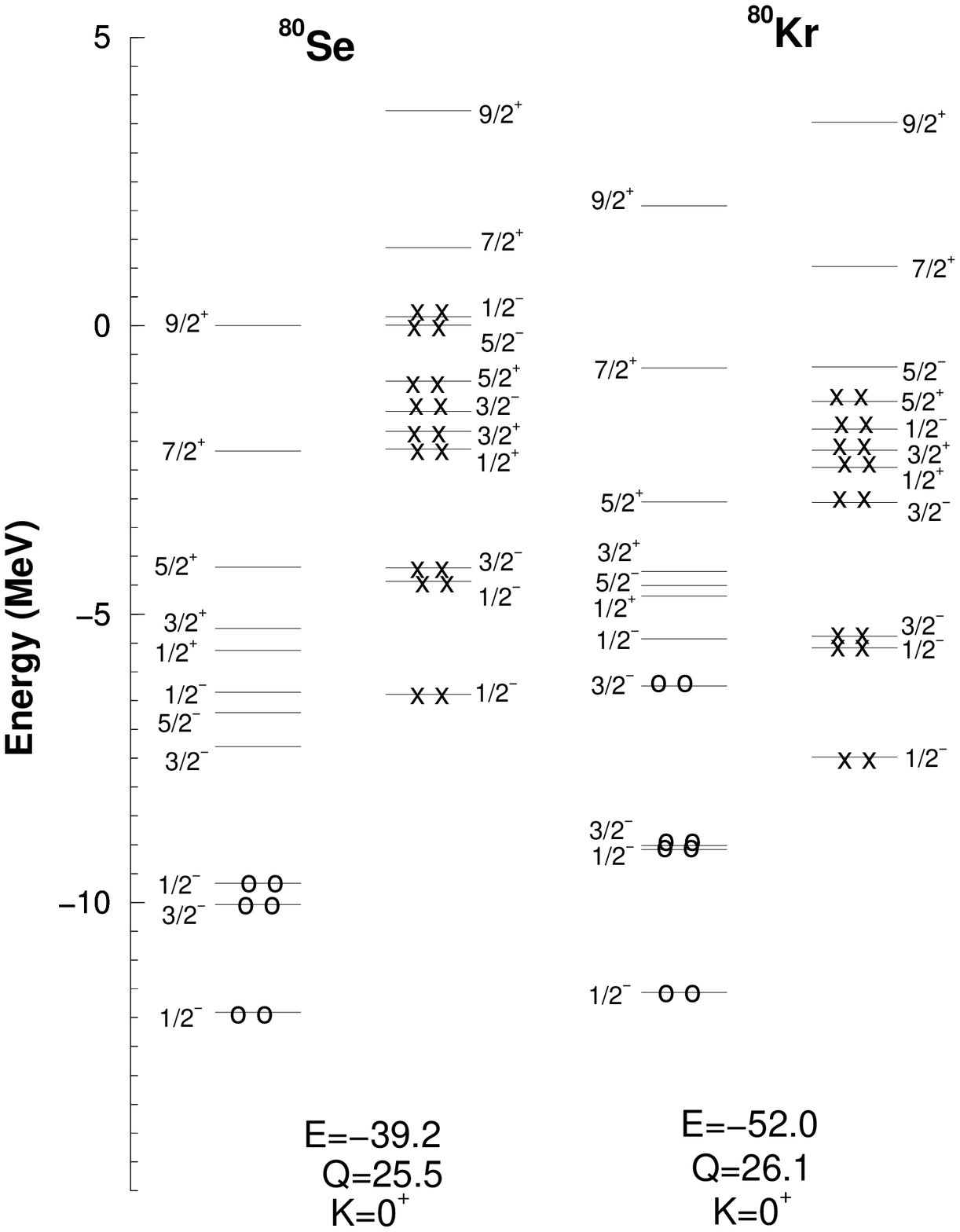}
\caption{
HF single particle spectra for $^{80}$Se and  $^{80}$Kr
corresponding to lowest prolate configurations.  In the
figures circles represent protons and crosses represent neutrons. The
Hartree-Fock energy  ($E$) in MeV, mass quadrupole moment ($Q$) in units of
the square of the oscillator length parameter and the total $K$ quantum
number of the lowest intrinsic states are given in the figure. Each
occupied single particle orbital is two fold degenerate because of time
reversal symmetry.
}
\label{fig5}
\end{figure}

\begin{figure}
\includegraphics{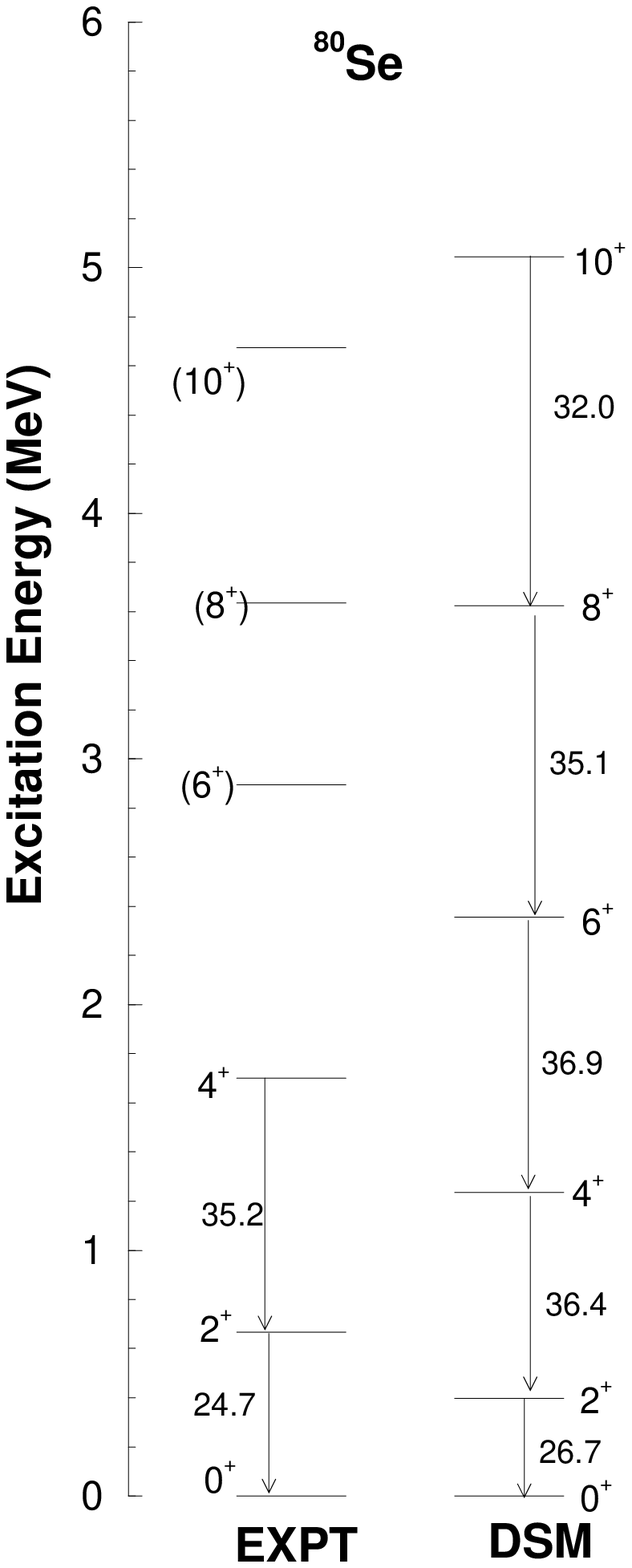}
\caption{
The calculated energy levels for $^{80}$Se are compared with
experiment. The experimental data are from ref \cite{nndc}.
The quantities near the arrows represent B(E2) values in W.u.
}
\label{fig6}
\end{figure}

\begin{figure}
\includegraphics[width=6in]{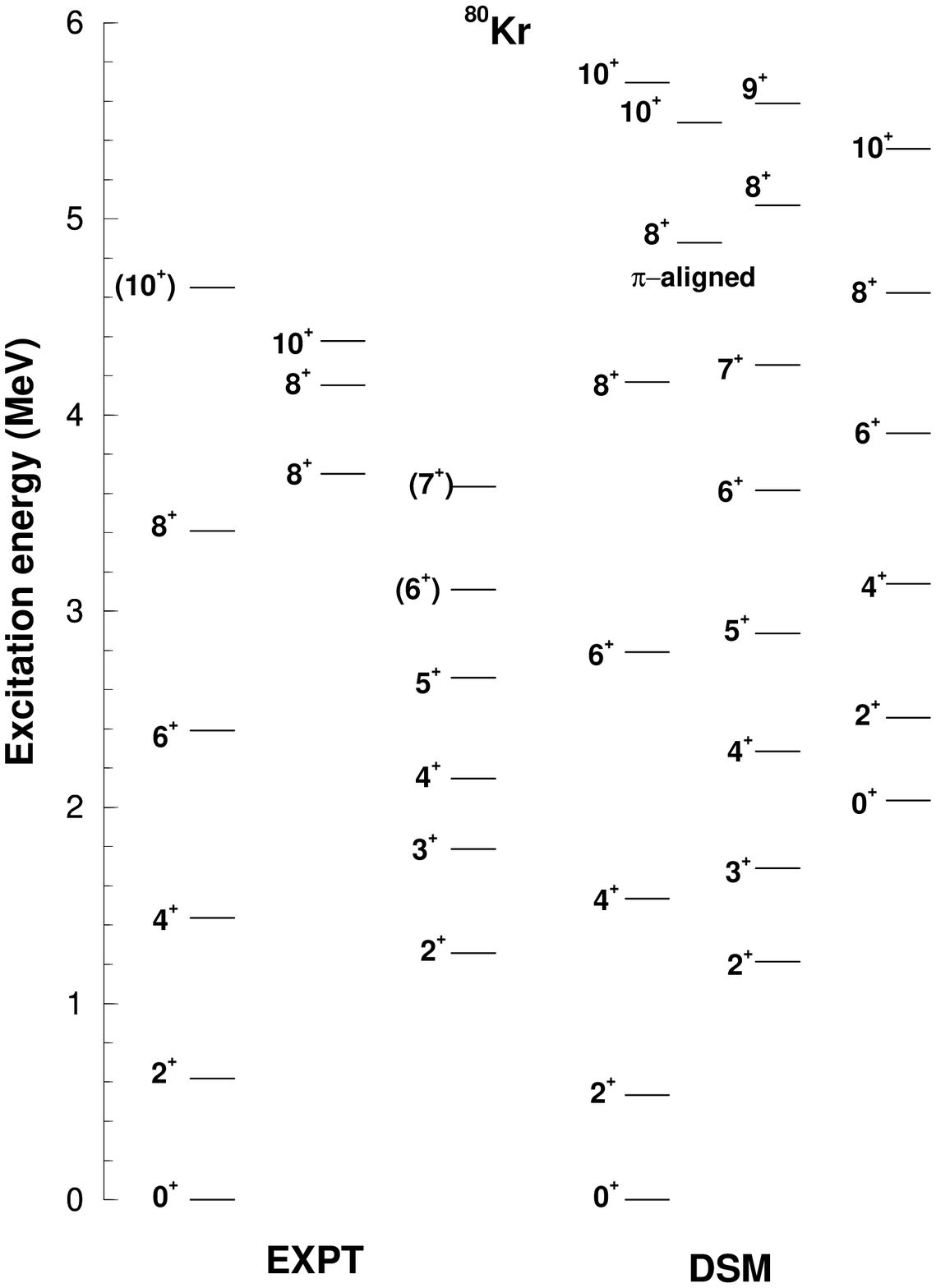}
\caption{
The calculated energy levels for $^{80}$Kr are compared with
experiment. The experimental data are from ref \cite{nndc}.
}
\label{fig7}
\end{figure}

In the double beta decay of $^{80}$Se, the daughter nucleus is $^{80}$Kr. As in
the case of $^{70}$Zn, we will first consider spectroscopic  properties of these
nuclei before discussing the double beta decay results. Both the nuclei are
studied using the modified  Kuo effective interaction given in \cite{Kuo} in the
model space  consisting of the single particle orbitals $^{2}p_{3/2}$,
$^{1}f_{5/2}$, $^{2}p_{1/2}$ and  $^1g_{9/2}$ with $^{56}$Ni as the core. This
effective interaction was used by us in most of our calculations for A=60-90
nuclei. The spherical single particle energies for these orbitals are taken as
0.0, 0.78, 1.08 and 4.25 MeV for protons and 0.0, 0.78, 1.58 and 2.75 MeV for
neutrons.  It may be mentioned that we had performed a preliminary study  of the
spectroscopy of $^{80}$Kr and $^{82}$Kr \cite{Kct1} with  spherical single
particle energies of neutrons to be same as protons taken above.  However the
present set of single particle energies gives much better description of
spectroscopic properties.  The HF single particle spectrum for these two nuclei
are given in Fig. \ref{fig5}. For  $^{80}$Se, there is a well defined gap of 2.3
MeV above the proton Fermi surface and the gap above the neutron Fermi surface
is 1.2 MeV. Thus,  $^{80}$Se has a relatively stable shape. On the other hand,
for  $^{80}$Kr, the proton and neutron gaps above their respective Fermi
surfaces are much smaller. For studying spectroscopic properties of  $^{80}$Se,
we have considered 10 intrinsic states (including the lowest one). The excited
configurations are obtained by making particle-hole excitations over the lowest
configuration. As discussed above, good angular momentum states are projected
from each of these intrinsic states and then these good $J$ states are
orthonormalized. The calculated spectrum is compared with experiment in Fig.
\ref{fig6}.  The agreement is quite satisfactory for the ground band. The B(E2)
values are given near the arrows in the figure and the known B(E2)s are well
reproduced. In order to study the energy spectrum of the daughter nucleus
$^{80}$Kr, we have considered a total of 12 intrinsic states, the excited ones
are obtained by particle-hole excitation. The energy spectrum obtained after
angular momentum projection and band mixing is shown in Fig. \ref{fig7}.
Experimentally for this nucleus, the ground band up to $J=10^+$  and a
quasi-gamma band have been observed. This nucleus shows three $8^+$ and two
$10^+$ levels. The ground band and the quasi-gamma band are quite well 
reproduced in our calculation. We find that a proton aligned band crosses the
ground band at $J=10^+$. In addition, we predict an excited $K=0^+$ band. In our
calculation, we also obtain three close lying $8^+$ and two close lying $10^+$
levels as in the experiment. We have calculated B(E2) values for all possible
transitions. The comparison with experiment for some of the B(E2) values are
given in Table \ref{tab2}.  In the calculation of B(E2), we have used the
effective charges 1.6 and 1.0 for protons and neutrons as in our earlier
calculations with this effective interaction.  For both the nuclei, we  have
also calculated the orbit occupancies and they are displayed in Fig. 
\ref{fig4}.

\subsection{$2\nu$ DBD half lives and $0\nu$ DBD NTME and half lives}

The $2\nu \beta \beta$ decay for this nucleus is studied using the Eqns.
\ref{eqn19} and \ref{eqn20}. The atomic masses of $^{80}$Se, $^{80}$Kr and the
intermediate nucleus $^{80}$Br are -77.7599, -77.8925 and -75.8895 MeV
respectively as given in \cite{audi-1}. For the intermediate nucleus $^{80}$Br,
$1^+$ is the ground state. Taking $E_{1^+}$ to be the calculated energies of
$^{80}$Br with respect to the lowest $1^+$, the energy denominator  is given by
$[ E_N - (E_I + E_F)/2 ] = [1.9367 + E_{1^+}]$ MeV.   DSM is used to calculate
$E_{1^+}$. In the DSM calculation for  $2\nu \beta \beta$, we have considered 13
configurations for $^{80}$Se, 55 configurations for the daughter nucleus
$^{80}$Kr, all with $K=0^+$. For the intermediate nucleus $^{80}$Br, we
considered 99 configurations with $K=1^+$ or $K=0^+$. For the odd-odd nucleus
like $^{80}$Br, the configurations with $K=0^+$ can also give $1^+$ levels. We
project out $0^+$ levels for the parent and daughter nuclei and then
orthonormalize them separately. Similarly, for the intermediate nucleus
$^{80}$Br, we project out $1^+$ levels from all the configurations and then
orthonormalize them. The calculated half life with phase space factor $0.12
\times 10^{-27}$ \cite{Vogel} comes out to be $1.97 \times  10^{29}$ yr. This
value agrees quite well with the QRPA result quoted by Bobyk et al.
\cite{Bobyk}. The nuclear matrix element $M_{2\nu}$ is 0.40 $MeV^{-1}$. 
The contribution to the nuclear matrix element from the first two $1^+$ states
of the intermediate nucleus is 0.038 and 0.288 $MeV^{-1}$. The contribution
of other $1^+$ states are more than 10 times smaller. Thus, this nucleus 
shows the single-state dominance in the $2\nu\beta\beta$ decay transition
as predicted in \cite{CS-1}.

Then we proceed to calculate the nuclear matrix element for neutrinoless double
beta decay. As above we considered 13 configurations for the parent nucleus 
$^{80}$Se and 55 intrinsic states for the daughter nucleus $^{80}$Kr. The $0\nu
\beta\beta$ nuclear matrix element comes out to be 3.19.  The GT and the Fermi
matrix elements contributing to the above NTME are found  to be 2.684 and
-0.801, respectively. Here also we find that the Fermi matrix element is
substantially smaller than the GT matrix element indicating that there is no
isospin contamination. Also, to our knowledge there are no other model
calculations available for $M^{0\nu}$ for $^{80}$Se. Going further, taking the
neutrino mass to be 1 eV and phase space factor  $0.43 \times 10^{-28}$
yr$^{-1}$ \cite{Vogel}, the half-life is  $2.3 \times 10^{27}$ yr. The
half-lives are displayed in table \ref{tab3}. 

\section{$^{82}$Se results}

\subsection{Spectroscopic properties}

\begin{figure}
\includegraphics[width=6in]{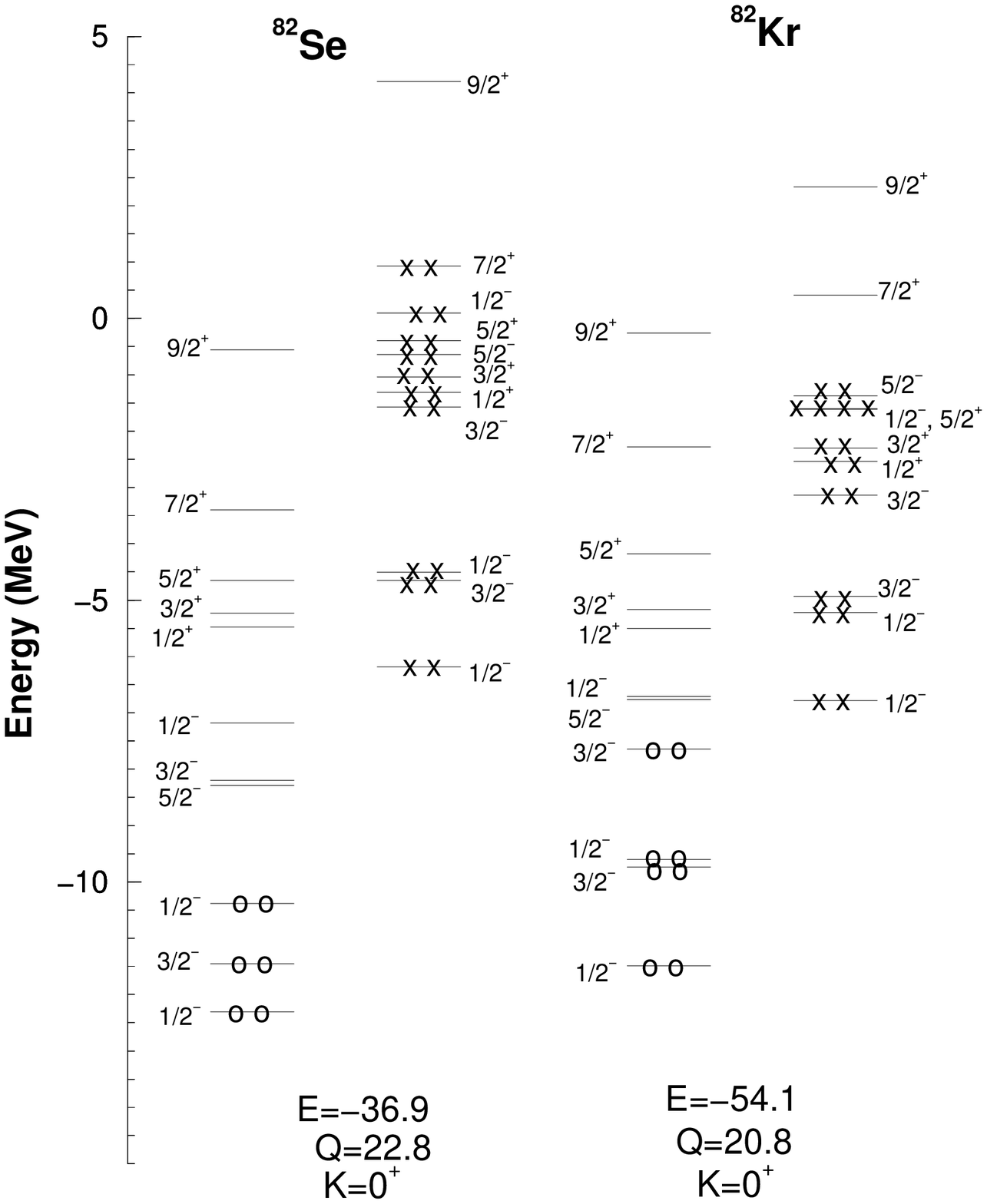}
\caption{
HF single particle spectra for $^{82}$Se and  $^{82}$Kr
corresponding to lowest prolate configurations.  In the
figures circles represent protons and crosses represent neutrons. The
Hartree-Fock energy  ($E$) in MeV, mass quadrupole moment ($Q$) in units of
the square of the oscillator length parameter and the total $K$ quantum
number of the lowest intrinsic states are given in the figure. Each
occupied single particle orbital is two fold degenerate because of time
reversal symmetry.
}
\label{fig8}
\end{figure}

\begin{figure}
\includegraphics{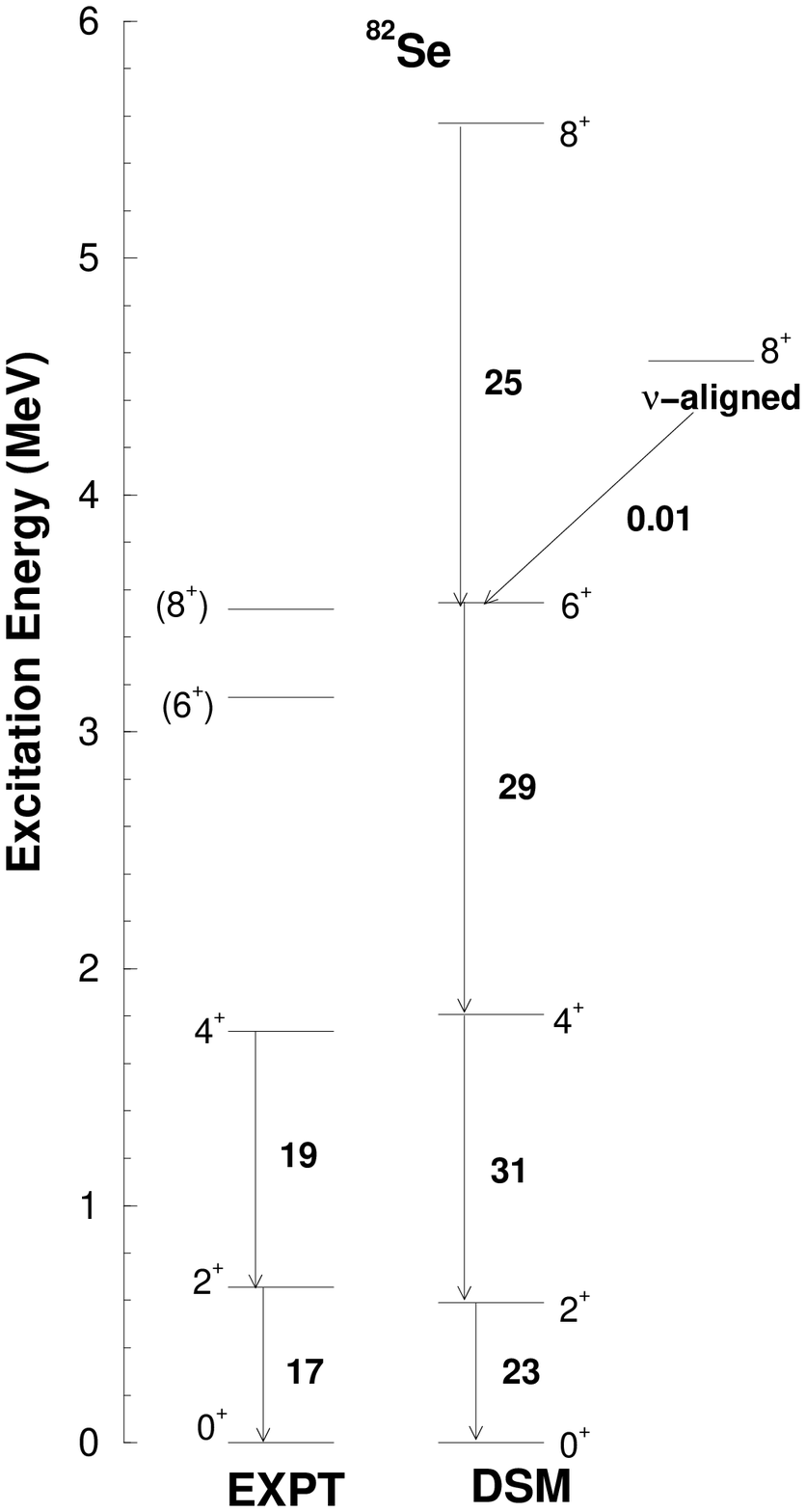}
\caption{
The calculated energy levels for $^{82}$Se are compared with
experiment. The experimental data are from ref \cite{nndc}.
}
\label{fig9}
\end{figure}

As in the case of $^{80}$Se, we perform the calculation of $^{82}$Se and the
daughter nucleus  $^{82}$Kr using the modified Kuo effective interaction in the
model space consisting of  the single particle orbitals $^{2}p_{3/2}$,
$^{1}f_{5/2}$, $^{2}p_{1/2}$ and $^1g_{9/2}$ with $^{56}$Ni as the core.  The
spherical single particle energies for these orbitals are taken as 0.0, 0.78,
1.08 and 4.25 MeV for protons and 0.0, 0.78, 1.58  and 2.75 MeV for neutrons. 
As mentioned earlier, the same model space was used in many shell model and 
IBM calculations \cite{sm-ik1,sm-ik2,sm-ik3,sm-ik4,sm-ik5,IBM2}. 
As before, we first perform axially symmetric HF calculation for  both the
nuclei and obtain the lowest energy prolate solution. The lowest  energy prolate
solution is shown in Fig. \ref{fig8}.  $^{82}$Se also has a well defined gap 2.1
MeV and 3.2 MeV above the proton and neutron Fermi  surfaces and hence it has a
stable shape. However for the daughter nucleus $^{82}$Kr, the gaps are much
smaller. The proton gap is only 0.9 MeV whereas the neutron gap is 1.8 MeV. For
calculating the energy spectrum, we perform particle-hole excitations over the
lowest HF configuration for each nucleus and obtain excited intrinsic states.
Good angular momentum states are projected from each of these intrinsic states.
These good angular momentum states are then orthonormalized and then a band
mixing calculation is performed. For studying the energy spectrum of $^{82}$Se,
we considered 10 intrinsic states. The energy spectrum is given in Fig.
\ref{fig9}. The ground band is quite well reproduced in our calculation. We find
that a neutron aligned band crosses the ground band at $J=8^+$. Recently we had
studied the $2\nu \beta \beta$ decay of this nucleus \cite{SSK-2} using a
slightly different set of spherical single particle energies. The energy
spectrum is more or less similar to the present case.  We have calculated B(E2)
values for all possible transitions taking effective charges 1.6 and 1.0 for
protons and neutrons. In the ground band, only two B(E2) transitions are know
and they are in good agreement with  DSM values as shown in Fig. 9. 

\begin{figure}
\includegraphics[width=6in]{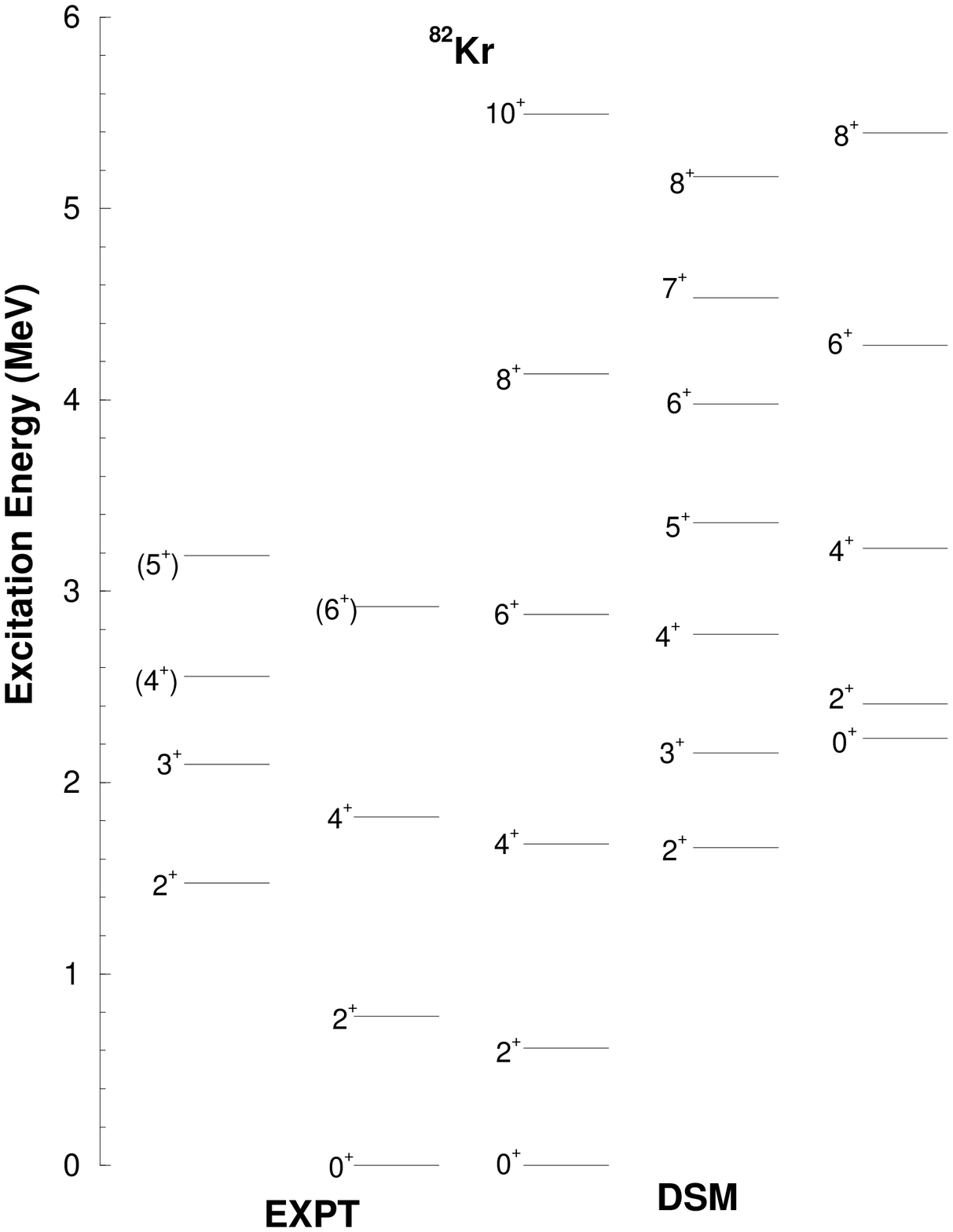}
\caption{
The calculated energy levels for $^{82}$Kr are compared with
experiment. The experimental data are from ref \cite{nndc}.
}
\label{fig10}
\end{figure}

For calculating the spectroscopic properties of the daughter nucleus $^{82}$Kr,
we have considered 12 intrinsic states. After angular momentum projection and
orthonormalization, the energy bands obtained in our calculation are  compared
in Fig. \ref{fig10}. The ground band and the quasi-gamma band are quite nicely
reproduced in our calculation. We calculate the B(E2) values with the same
effective charges as in $^{80}$Se. The B(E2) values are compared with experiment
in table \ref{tab2}. The agreement is quite  satisfactory.

\subsection{$2\nu$ DBD half lives and $0\nu$ DBD NTME and half lives}

The $2\nu$ DBD for $^{82}$Se is first studied using Eqns. \ref{eqn19} and
\ref{eqn20}. We have considered 7 intrinsic states for $^{82}$Se and 48
intrinsic states for the daughter nucleus $^{80}$Kr all with $K = 0^+$. $J =0^+$
states are projected out from each of these intrinsic states and then these
states are orthonormalized for each nucleus. The intermediate nucleus $^{82}$Br
generates the $1^+$ states.  Excited configurations with $K=0^+$ and $K=1^+$ for
the intermediate  nucleus are obtained by making particle-hole  excitations over
the lowest HF intrinsic configuration. Then we  perform angular momentum
projection to project out $J=1^+$ levels from each intrinsic state. These 
angular momentum states are then orthonormalized. We took 83 configurations with
$K=0^+$ or $K=1^+$ and calculated the numerator of Eq. (\ref{eqn20}). The atomic
masses for the parent, intermediate and daughter nuclei are -77.5940, -77.4965
and -80.5895 MeV respectively  \cite{audi-1}. For this nucleus, the lowest $1^+$
state is at an excitation of 0.075 MeV.  If $E_{1^+}$ are the energies of the
excited $1^+$ states with respect to the lowest one which are calculated within
our DSM model, the energy denominator is given by (1.6702 + $E_{1^+}$). With the
phase space factor $0.43 \times 10^{-17}$ yr$^{-1}$ \cite{Vogel}, we  predict
the half life to be $1.58 \times 10^{19}$ yr compared to the  experimental value
$(0.92 \pm 0.07) \times 10^{20}$ yr. Thus our calculated value is about 6 times
smaller than experiment. Similar effect was seen by Caurier et al 
\cite{Caurier-12} in their shell model calculation and hence they have to 
introduce a quenching factor of 0.60. If we use this factor, our half life
is close to the shell model value. The nuclear matrix element is 0.24 
$MeV^{-1}$.
The contribution of the first $1^+$ state of the intermediate nucleus to the
nuclear matrix element is 0.21 $MeV^{-1}$ which is predominant. The contribution
of the other $1^+$ states to the nuclear matrix element is more than 10
times smaller. Thus this nucleus also exhibits single-state dominance in
$2\nu\beta\beta$ decay transitions as predicted in ref. \cite{CS-1}.

Then we proceed to calculate the neutrinoless double beta decay half life.  We
take 7 configurations for the parent nucleus $^{82}$Se and 48 configurations
for  the daughter nucleus $^{82}$Kr as in the two neutrino case with $K = 0^+$.
Good $J=0^+$ states are projected from these intrinsic states and they are
orthonormalized by performing band mixing calculation for each nucleus. The
calculated  nuclear matrix element is 2.04. 
The GT contribution is found to be 1.853 and the Fermi matrix element whose
magnitude is
smaller than the GT matrix element is -0.296. Taking the
phase  space factor $0.11 \times 10^{-24}$ yr$^{-1}$ \cite{Vogel} and the 
neutrino mass to be 1eV, the half life is  $2.2 \times 10^{24}$ yr. The half
lives are presented in table \ref{tab3}. It should be added that there is
considerable interest in $^{82}$Se because of the upcoming SuperNEMO experiment
\cite{Se-1}. The NEMO-3 gave a lower limit of $3.2 \times 10^{23}$ yr at 90\%
C.L. for this nucleus \cite{Se-1}. 

With SuperNEMO coming soon, new large scale shell model calculations are being
carried out \cite{sm-ik2}. There are also predictions for $M^{0\nu}$ from many
other nuclear models as shown in Table II and the results are as follows. Shell
model values range from 2.18 to 3.39 \cite{SM,sm-ik2,Men-2009} and the  current
expectation is that its value should be $3.3 \pm 0.1$ as quoted in 
\cite{sm-ik2}. The DSM value as given above is 2.04 and it is smaller than the
SM value. Similarly, various QRPA calculations give $M^{0\nu}$ in the range 2.77
to 5.65 \cite{QRPA3,JS1,Si-09,Rod-1}. The current value from QRPA with Jastrow
short range correlations is $M^{0\nu}=3.15 \pm 0.3$ and from UCOM correlations
is $4.2 \pm 0.35$ \cite{JS1}. Finally the values from IBM-2 model are 4.41 to
4.84 \cite{IBM1,IBM2,sm-ik5} and from GCM+PNAMP model is 4.22 \cite{DFT}.

\section{Conclusions}

In the A=60-90 region with $30 \leq \mbox{Z} \leq 40$ and $\mbox{N} \leq 48$
there are eight candidate nuclei, as shown in Table I, which can undergo double
beta decay.  Prompted by the success of deformed shell model, based on
Hartree-Fock intrinsic states with band mixing, in explaining spectroscopic
properties of nuclei in the A=60-90 region, recently DSM results for the four 
$0\nu\,e^+$DBD candidate nuclei, i.e. $^{64}$Zn, $^{74}$Se, $^{78}$Kr and
$^{84}$Sr have been reported by us in \cite{SSK}. Similarly the results for
$2\nu\;e^+$DBD half lives are reported in \cite{SK-kr,SK-se,SK-sr,SSK}. In
addition, results for $2\nu$ and $0\nu$\,DBD for $^{76}$Ge have been reported in
\cite{ge76} and \cite{SSK} respectively. Complementing this work, in the 
present paper DSM results for $0\nu$\,DBD for the three remaining nuclei, i.e.
for $^{70}$Zn, $^{80}$Se and $^{82}$Se are presented. After ensuring reasonable
DSM description of spectroscopic properties that include spectra and $B(E2)$'s,
we have presented the results for occupancies for the parent and daughter nuclei
involved. Proceeding further, DSM results for NTME and there by for the
$0\nu\,$DBD half-lives are presented for $^{70}$Zn, $^{80}$Se and $^{82}$Se as
shown in Table II.  The present paper brings to conclusion  DSM study of the
eight DBD candidate nuclei listed in Table I. Future DSM studies call for using 
much larger set of single particle levels, suitable effective interactions in
the larger spaces and inclusion of much larger number of intrinsic states in
band mixing calculations.  The present model space has also been used in many
shell model \cite{sm-ik1,sm-ik2,sm-ik3,sm-ik4} and interacting boson model
studies  \cite{IBM2,sm-ik5}.  We plan to expand the model space in future to
include much larger number of spin-orbit partners so that the Ikeda sum rule is
well satisfied and also carry out an analysis of (p,n) GT strength functions and
beta decay data involving $0\nu$ and $2\nu$ DBD nuclei.

Before concluding,  we mention that recently we have also studied the role of 
deformation in generating NTME for $0\nu$ DBD within DSM framework.  Towards
this end we have considered $^{70}$Zn and $^{150}$Nd nuclei as examples and
employed for each of these two different effective interactions that produce 
spherical and well deformed shape respectively. These calculations clearly
showed that  deformation reduces NTME by a factor of 2-3; see \cite{SK-2014} for
details. In a different approach, Men\'{e}ndez et al \cite{Men-2009} examined
pairing effects (where the shape is close to spherical) 
versus deformation within
shell model by analyzing the evolution of NTME with the maximum seniority 
permitted in the wavefunctions.

\begin{acknowledgments}

\noindent 

RS is thankful to DST (Government of India) for financial support. RS and VKBK
thank H.J. Kim, KNU, South Korea for his interest in the present work.
Thanks are also to the referee for many useful comments.

\end{acknowledgments}

\newpage

\begin{table}
\caption{DSM results for half-lives for $0\nu\;\beta\beta$ with $m_\nu = 1$ eV
(there are claims that the effective neutrino mass is certainly much less than
$1$ eV \cite{planck1,planck2}). The values of $G^{0\nu}$ given in column 2
are taken from Ref. \cite{Vogel}. Columns 3 and 4 are DSM results.
Fifth column gives current experimental bounds for the half lives. 
The last column gives the values of NTME predicted by other models
along with the references. Note that in the table QRPA$^{1)}$ corresponds to
calculations using QRPA with $g_A=1$ and QRPA$^{2)}$ corresponds to $g_A=1.25$.
See text for further details.}
\begin{tabular}{ccccccc}
\hline \hline
Nucleus  & $G^{0\nu}$(yr$^{-1}$)  & $M^{0\nu}$ & $T_{1/2}$ (yr) & 
$T_{1/2}$ (yr) & $M^{0\nu}$ \\
& & (DSM) & (DSM) & Expt'l bound (yr) & (other models) \\
\hline
$^{70}$Zn & $0.23 \times 10^{-26}$  & 1.99 & $1.1 \times 10^{26}$ & $\geq 3.2
\times 10^{19}$ & $2.93-3.58$ QRPA$^{1)}$ \cite{Suh-3} \\
& & & & & $1.44-2.47$ QRPA$^{2)}$ \cite{Suh-3} \\
$^{80}$Se & $0.43 \times 10^{-28}$  & 3.19& $2.3 \times 10^{27}$ &  & \\
$^{82}$Se&  $0.11 \times 10^{-24}$  & 2.04 & $2.2 \times 10^{24}$ & $\geq 
3.2 \times 10^{23}$ & $2.18-3.39$ (SM) \cite{SM,sm-ik2,Men-2009} \\
& & & & & $2.77-5.65$ (QRPA) \cite{QRPA3,JS1,Si-09,Rod-1}\\
& & & & & $4.41-4.84$ (IBM-2) \cite{IBM1,IBM2,sm-ik5} \\
& & & & & $4.22$ (GCM+PNAMP) \cite{DFT} \\
\hline
\hline
\label{tab3}
\end{tabular}
\end{table}

\newpage

\begin{table}
\caption{DSM model predicted B(E2;$J_i \rightarrow J_f$) values in W.u. for
$^{80}$Kr and $^{82}$Kr are compared with experimental data given in
\cite{nndc}.}
\begin{tabular}{cccccccc}
\hline \hline
\multicolumn{4}{c}{B(E2)'s for $^{80}$Kr} & \multicolumn{4}{c}{B(E2)'s for $^{82}$Kr}\\
\hline
$J_i$  & $J_f$ & DSM   & Expt.           & $J_i$ & $J_f$ & DSM & Expt.\\
\hline

$2^+$  & $0^+$ & 23.1  & 37.3 $\pm$ 2.2  & $2^+$  & $0^+$ & 17.8 & 21.3 $\pm$ 0.7 \\
$2'^+$ & $2^+$ & 9.2 & 25 $\pm$ 5        & $2'^+$  & $2^+$ & 8.4 & $\approx$ 5.5 \\
$2'^+$ & $0^+$ & 2.1 & 0.30 $\pm$ 0.07   & $2'^+$  & $0^+$ & 2.6 &  \\
$4^+$ & $2^+$ & 33.9 & 70 $\pm$10        & $3'^+$  & $2^+$ & 4.6  &           \\
$3'^+$& $2'^+$ &38.4 & 34 $\pm$ 5        & $3'^+$  & $2'^+$ & 30.1  & \\
$3'^+$& $2^+$  & 3.6 & 0.57 $\pm$ 0.14   & $4^+$   & $2^+$ & 25.5 & 32 $\pm$ 12  \\
$4'^+$& $4^+$  & 7.2 &32 $\pm$ 20       & $4'^+$  & $2'^+$ & 6.0 & 9 $\pm$ 3 \\
$4'^+$& $2'^+$  & 11.0 & 50 $\pm$ 30     & $6^+$  & $4^+$ & 27.9 & 5.5 $\pm$ 1.9  \\
$4'^+$& $2^+$  & 0.1  & 0.26 $\pm$ 0.18  & $5'^+$ &  $4^+$ & 0.3 & 7.3 $\pm$ 2.1  \\
$6^+$ & $4^+$  & 38.8 & 62 $\pm$ 16      & $8^+$ & $6^+$  & 27.4 &    \\
$5'^+$ & $3'^+$  & 19.5 & 50 $\pm$ 17   & $10^+$ & $8^+$  & 25.6&                \\
$5'^+$ & $4^+$  & 1.3 & 1.2 $\pm$ 0.7   &        &        &        &   \\
$6'^+$ & $6^+$  & 4.1  & 17 $\pm$ 15    &        &        &        &         \\
$6'^+$ & $4'^+$  & 20.2 & 33 $\pm$ 17   &        &        &        &   \\
$6'^+$ & $4^+$  & 0.1   & $<$0.23       &        &        &        &   \\
$8^+$ & $6^+$  & 38.7   & 90$ ^{+90}_{-45}$  &   &        &        &    \\
$10^+$& $8^+$  & 11.1 &                 &        &        &        &    \\
\hline
\hline
\label{tab2}
\end{tabular}
\end{table}

\newpage

\newpage

\newpage

\newpage

\newpage

\newpage

\newpage

\newpage

\newpage

\newpage

\begin{thebibliography}{99}

\bibitem{Kam} A. Gando {\it et al.}, Phys. Rev. Lett. {\bf 110}, 062502 (2013).

\bibitem{Exo} M. Auger {\it et al.}, Phys. Rev. Lett. {\bf 109}, 032505 (2012).

\bibitem{Gerda} M. Agostini {\it et al.}, Phys. Rev. Lett. {\bf 111}, 122503
(2013).

\bibitem{SM} E. Caurier, J. Men\'{e}ndez, F. Nowacki, and A. Poves, Phys. Rev. 
Lett. {\bf 100}, 052503 (2008)

\bibitem{QRPA1} F. \v{S}imkovic, A. Faessler, V. Rodin, P. Vogel and J. Engel,
Phys. Rev. C {\bf 77}, 045503 (2008)

\bibitem{QRPA2} D.-L. Fang, A. Faessler, V. Rodin and F.\v{S}imkovic,
Phys. Rev. C {\bf 83}, 034320 (2011)

\bibitem{QRPA3} M. Kortelainen and J. Suhonen, Phys. Rev. C {\bf 75}, 051303(R)
(2007).

\bibitem{IBM1} J. Barea, J. Kotila, and F. Iachello, Phys. Rev. Lett. 
{\bf 109}, 042501 (2012).

\bibitem{DFT} T.R. Rodr\'{i}guez and G. Martinez-Pinedo, Phys. Rev. Lett. 
{\bf 105}, 252503 (2010)

\bibitem{HFB} K. Chaturvedi, R. Chandra, P.K. Rath, P.K. Raina, and J.G. Hirsch,
Phys. Rev. C {\bf 78}, 054302 (2008)

\bibitem{IBM2} J. Barea, J. Kotila, and F. Iachello, Phys. Rev. C 
{\bf 87}, 014315 (2013).

\bibitem{JS1} J. Suhonen and O. Civitarese, J. Phys. G {\bf 39}, 124005 (2012).

\bibitem{JS2} J. Suhonen, Eur. Phys. J. A {\bf 48}, 51 (2012)

\bibitem{zn64} B P. Belli {\it et al.},  J. Phys. G: Nucl. Part. Phys. {\bf 38}, 115107 (2011).

\bibitem{se74} A.S. Barabash, P. Hubert, A. Nachab, and V. Umatov, 
Nucl. Phys. A {\bf 785}, 371 (2007).

\bibitem{sr84} \url{
http://www.apctp.org/topical/2009/npap2009/Presentations/APCTP_NPAP2009_192_Kim.pdf}

\bibitem{sahu1} R. Sahu and S.P. Pandya, J. Phys. G {\bf 14}, L165 (1988).

\bibitem{sahu2} R. Sahu and S.P. Pandya, Nucl. Phys. {\bf A548}, 64 (1992).

\bibitem{sahu3} K.C. Tripathy and R. Sahu, J. Phys. G {\bf 20}, 911 (1994).

\bibitem{br8082} R. Sahu and S.P. Pandya, Nucl. Phys. {\bf A571}, 253
(1994).

\bibitem{npa96} K. C. Tripathy and R. Sahu, Nucl. Phys. {\bf A597}, 177
(1996).

\bibitem{sahu4} K.C. Tripathy and R. Sahu, Int. J. Mod. Phys. E {\bf 11},
531 (2002).

\bibitem{sk1} R. Sahu and V.K.B. Kota, Phys. Rev. {\bf C 66}, 024301
(2002).

\bibitem{sk2} R. Sahu and V.K.B. Kota, Phys. Rev. {\bf C 67}, 054323 (2003).

\bibitem{msk} S. Mishra, R. Sahu, and V.K.B. Kota, Prog. Theo. Phys. {\bf
118}, 59 (2007).

\bibitem{app1}  T.S. Kosmas, A. Faessler, and R. Sahu, Phys. Rev. C
{\bf 68}, 054315 (2003).

\bibitem{app3} R. Sahu, K.H. Bhatt, and D.P. Ahalpara, J. Phys. G {\bf 16},
733 (1990).

\bibitem{ge76} R. Sahu, F. \v{S}imkovic, and A. Faessler, J. Phys. G
{\bf25}, 1159 (1999).

\bibitem{ga62} P.C. Srivastava, R. Sahu and V.K.B. Kota, Eur. Phys. J. A {\bf
51}: 3 (2015).

\bibitem{SK-kr}  S. Mishra, A. Shukla, R. Sahu, and V.K.B. Kota, Phys. Rev. C
{\bf 78}, 024307 (2008).

\bibitem{SK-se} A. Shukla, R. Sahu and V.K.B. Kota, Phys. Rev. C {\bf 80},
057305 (2009).

\bibitem{SK-sr} R. Sahu and V.K.B. Kota, Int. J. Mod. Phys. E {\bf 20},
1723 (2011).

\bibitem{SSK} R. Sahu, P.C. Srivastava and V.K.B. Kota, J. Phys. G {\bf 40},
095107 (2013)

\bibitem{SSK-2} R. Sahu, P.C. Srivastava and V.K.B. Kota, Can. J. Phys. 
{\bf 89}, 1101 (2011)

\bibitem{IBM3} J. Kotila and F. Iachello, Phys. Rev. C {\bf 87}, 024313 (2013).

\bibitem{En-88} J. Engel, P. Vogel and M. R. Zirnbauer, Phys. Rev. C {\bf
37}, 731-746 (1988)

\bibitem{To-91} T. Tomoda, Rep. Prog. Phys. {\bf 54}, 53-126 (1991).

\bibitem{Si-09}  F. \v{S}imkovic, A. Faessler, H. M\"{u}ther, V. Rodin and
M. Stauf, Phys. Rev. C {\bf 79}, 055501 (2009).

\bibitem{Ho-10} M. Horoi and S. Stoica, Phys. Rev. C {\bf 81}, 024321
(2010).

\bibitem{Kort-07} M. Kortelainen, O. Civitarese, J. Suhonen and J. Toivanen,
Phys. Lett. B {\bf 647}, 128 (2007).

\bibitem{Brown} B.A. Brown and A.F. Lisetskiy (unpublished);
See endnote (Ref. [28]) in B. Cheal et al. Phys. Rev. Lett. {\bf 104}, 252502
(2010).

\bibitem{sm-ik1} R.A. Senkova and M. Horoi, Phys. Rev. C {\bf 90},
051301 (2014).

\bibitem{sm-ik2} R.A. Senkov, M. Horoi and B.A. Brown, Phys. Rev. C {\bf 89},
054304 (2014).

\bibitem{sm-ik3} D.L. Lincoln et al, Phys. Rev. Lett. {\bf 110}, 012501 (2013).

\bibitem{sm-ik4} J. Menendez, A. Poves, E. Caurier and F. Nowacki,
Phys. Rev. C {\bf 80}, 048501 (2009).

\bibitem{sm-ik5} J. Barea and F. Iachelo, Phys. Rev. C {\bf 79}, 044301 
(2009).

\bibitem{nndc} ENSDF Data Base, Brookhaven National Laboratory, USA,
\url{http://www.nndc.bnl.gov/ensdf/index.jsp}.

\bibitem{Sch1} D. von Ehrenstein and J.P. Schiffer, Phys. Rev. {\bf 164}, 1374
(1967).

\bibitem{Ko-79} V.K.B. Kota and V. Potbhare, Nucl. Phys. {\bf A331}, 93 (1979). 

\bibitem{Sch2} J.P. Schiffer et. al., Phys. Rev. Lett. {\bf 108}, 022501 (2012).

\bibitem{Sch3} J.P. Schiffer {\it et al.} Phys. Rev. Lett. {\bf 100}, 112501
(2008).

\bibitem{Sch4} B.P. Kay et. al. Phys. Rev. C {\bf 79}, 021301(R) (2009).
\bibitem{Vogel} F. Boehm and P. Vogel, {\it Physics of Massive Neutrinos}
(Cambridge University Press, Cambridge, 1992).

\bibitem{audi-1} G. Audi, A.H. Wapstra and C. Thibault, Nucl. Phys. A{\bf 729},
337 (2003).

\bibitem{Bobyk} A. Bobyk et al. Nucl. Phys. A {\bf 669}, 221 (2000)

\bibitem{Suh-3} J. Suhonen, Nucl. Phys. A {\bf 864} 63 (2011).

\bibitem{CS-1} O. Civitarese and J. Suhonen, Nucl. Phys. A {\bf 653}, 321 (1999)

\bibitem{Kuo} D.P. Ahalpara, K.H. Bhatt and R. Sahu, J. Phys. G {\bf 11}, 
735 (1985)

\bibitem{Kct1} S. Mishra, K.C. Tripathy and R. Sahu, Can. J. Phys. {\bf 85},
269 (2007).

\bibitem{Caurier-12} Caurier et al Phys. Lett. B{\bf 711}, 62 (2012)

\bibitem{Se-1} L. Simard J. Phys. Conf. Ser. {\bf 375}, 042011 (2012).

\bibitem{planck1} P.A.R. Ade {\it et al.} (Planck collaboration), 
arXiv:1303.5076; Astronomy and Astrophysics {\bf 571}, A16 (2014).

\bibitem{planck2} P.S.B. Dev, S. Goswami, M. Mitra and W. Rodejohann, Phys. Rev.
D {\bf 88}, 091301 (2013).

\bibitem{Men-2009} J. Men\'{e}ndez, A. Poves, E. Caurier and P. Nowacki,
Nucl. Phys. A {818}, 139 (2009).

\bibitem{Rod-1} V.A. Rodin, A. Faessler, F. \v{S}imkovic, and P. Vogel, Nucl.
Phys. A {\bf 793}, 213 (2007)

\bibitem{SK-2014} R. Sahu and V.K.B. Kota, arXiv:1501.07674 [nucl-th] 30 Jan
2015.

\end{thebibliography}
\end{document}